\documentclass[letter,longauth,traditabstract]{aa} 
\usepackage{graphicx}
\usepackage{txfonts}
\usepackage{natbib}
\bibpunct{(}{)}{;}{a}{}{,}
\usepackage{subfigure}
\usepackage{lscape}

\begin{document}

   \title{The youngest massive protostars in the Large Magellanic Cloud\thanks{Herschel is an ESA space observatory with science instruments provided by European-led Principal Investigator consortia and with important participation from NASA.}
}

   \titlerunning{YSOs in the LMC}

    \author{M.~Sewi{\l}o\inst{1}
   \and R.~Indebetouw\inst{2}
   \and L.~R.~Carlson\inst{3}
   \and B.~A.~Whitney\inst{4}
   \and C.-H.~R.~Chen\inst{2}
   \and M.~Meixner
   \inst{1}\fnmsep\thanks{Visiting Scientist at Smithsonian Astrophysical Observatory, Harvard-CfA, 60 Garden St., Cambridge, MA, 02138, USA}
   \and T.~Robitaille\inst{5} \fnmsep\thanks{Spitzer Postdoctoral Fellow}
   \and J.~Th.~van~Loon\inst{6}
   \and J.~M.~Oliveira\inst{6}
   \and E.~Churchwell\inst{7}
   \and J.~D.~Simon\inst{8} 
   \and S.~Hony\inst{9}
   \and P.~Panuzzo\inst{9}
   \and M.~Sauvage\inst{9}
   \and J.~Roman-Duval\inst{1}
   \and K.~Gordon\inst{1}
   \and C.~Engelbracht\inst{10}
   \and K.~Misselt \inst{10}
   \and K.~Okumura\inst{9}
   \and T.~Beck\inst{1}
   \and J.~Hora\inst{5}
   \and P.~M.~Woods\inst{11}}

\institute{Space Telescope Science Institute, 3700 San Martin Drive, Baltimore, MD 21218, USA 
\email{mmsewilo@stsci.edu}
\and Department of Astronomy, University of Virginia, PO Box 3818, Charlottesville, VA 22903, USA 
\and Johns Hopkins University, Department of Physics and Astronomy, Homewood Campus, Baltimore, MD 21218, USA  
\and Space Science Institute,  4750 Walnut St. Suite 205, Boulder, CO 80301, USA 
\and Harvard-Smithsonian Center for Astrophysics, 60 Garden Street, Cambridge, MA, 02138, USA
\and School of Physical \& Geographical Sciences, Lennard-Jones Laboratories, Keele University, Staffordshire ST5 5BG, UK
\and Department of Astronomy, 475 North Charter St., University of Wisconsin, Madison, WI 53706, USA
\and Observatories of the Carnegie Institution of Washington,   813 Santa Barbara St., Pasadena, CA, 91101 USA
\and CEA, Laboratoire AIM, Irfu/SAp, Orme des Merisiers, F-91191 Gif-sur-Yvette, France 
\and Steward Observatory, University of Arizona, 933 North Cherry Ave., Tucson, AZ 85721, USA  
\and Jodrell Bank Centre for Astrophysics, Alan Turing Building, School of Physics \& Astronomy, University of Manchester, Oxford Road, Manchester M13 9PL, United Kingdom  
}

\date{Received ; accepted }

 
\abstract 
{We demonstrate the unique
capabilities of {\it Herschel} to study very young luminous extragalactic young stellar objects (YSOs) by analyzing a 
central strip of the Large  Magellanic Cloud obtained through the HERITAGE Science Demonstration Program.  
We combine PACS 100 and 160, and SPIRE 250, 350, and 500  $\mu$m photometry with 2MASS (1.25-2.17  $\mu$m) 
and {\it Spitzer} IRAC and MIPS  (3.6-70  $\mu$m) to construct complete spectral energy distributions (SEDs) of compact sources.
From these, we identify 207 candidate embedded YSOs in the observed region, $\sim$40\% never-before identified.
We discuss their position in far-infrared color-magnitude space, comparing with previously studied, 
spectroscopically confirmed YSOs and maser emission.  All have red colors indicating massive cool envelopes 
and great youth.
We analyze four example YSOs, determining their physical properties by fitting their SEDs with radiative 
transfer models.  Fitting full SEDs including the {\it Herschel} data requires us to increase the size and mass of 
envelopes included in the models.  This implies higher accretion rates ($\gtrsim$10$^{-4}$M$_\odot$yr$^{-1}$), 
in agreement with previous outflow studies of high-mass protostars.
Our results show that {\it Herschel} provides reliable longwave SEDs of large samples of high-mass YSOs; discovers 
the youngest YSOs whose SEDs peak in {\it Herschel} bands; and constrains the physical properties and evolutionary 
stages of YSOs more precisely than was previously possible.}


\maketitle


\section{Introduction}

The proximity of the Magellanic Clouds offers a unique opportunity to analyze the complete inventory of 
luminous YSOs over an entire galaxy.  With known YSO distances, luminosities, masses, and mass accretion 
rates can all be well-defined.  Comparison of the properties of YSOs in the  Magellanic 
Clouds and in the Milky Way can reveal differences in star formation physics due to 
metallicity and environment.  

Using the {\it Spitzer} SAGE (\textquotedblleft Surveying the Agents of Galaxy Evolution\textquotedblright) Survey of the Large Magellanic Cloud
 \citep[LMC;][]{meixner06}, 
Whitney et al. (2008; W08) and Gruendl \& Chu (2009; GC09) discovered $\sim$1800 massive YSO candidates 
in the LMC (a 90-fold increase over previous work). {\it Spitzer} studies selected sources using colors and SEDs at wavelengths $\leq$24$\;\mu$m (where {\it Spitzer} can resolve 
individual YSOs), requiring a detection at 4.5$\;\mu$m or shorter in most cases.  These surveys thus missed 
the youngest, most embedded YSOs that can only be detected at longer wavelengths.

The {\it Herschel} Space Observatory \citep{pilbratt10} has the spatial resolution required 
to study individual sources at $\lambda\gtrsim$50 $\mu$m (from $\sim$1.3 pc at 70 $\mu$m to $\sim$8.7 pc 
at 500 $\mu$m for a distance of 50 kpc, \citealt{schaefer2008}). 
The least-evolved massive protostars are characterized by cold dust temperatures probed at far-infrared (far-IR)  
wavelengths, and are expected to be $\sim$10$^3$ times brighter at 100 $\mu$m than at 5 $\mu$m 
\citep{whitney04,molinari08}, making {\it Herschel} extremely effective at detecting those youngest YSOs.

With {\it Herschel}, we not only discover new objects but also better characterize 
{\it Spitzer}-identified YSO candidates.  The {\it Herschel} data constrain the models of these sources and 
improve estimates of such physical parameters as total luminosity, stellar mass, and total dust mass.
We demonstrate these capabilities by studying a strip across the LMC
observed as part of the Science Demonstration Program (SDP) -- the
first part of the {\it Herschel} Key Program \textquotedblleft HERschel Inventory of
the Agents of Galaxy Evolution\textquotedblright\ (HERITAGE; \citealt{meixner10}) in the
Magellanic Clouds.  The strip was mapped in the PACS 100 and
160$\;\mu$m bands \citep{poglitsch10} and SPIRE 250, 350, and
500$\;\mu$m bands \citep{griffin10}.

\section{Source selection and photometry}  

An initial list of 640 sources was compiled by hand, choosing apparent point sources in {\it Herschel} images.  
Astrometry was refined using the SAGE-LMC MIPS 24$\;\mu$m image.
We performed aperture photometry on 2MASS, {\it Spitzer}, and {\it Herschel} images using apertures scaled to the instrumental 
resolution: 4$''$ radius for 2MASS, 3$''$ for IRAC, 6$''$ and 12$''$ for MIPS 24 and 70 $\mu$m, 8$''$ and 12$''$ 
for PACS 100 and 160, and 13$''$, 17$''$, and 23$''$ for SPIRE 250, 350, and 500, respectively.  
The aperture size relative to PSF was chosen after examination of the multi-wavelength images, to most 
consistently measure a single YSO as distinct from its environment.  Aperture corrections are 1.4, 1.5, 1.7, 
and 1.7 for IRAC, 1.8 for MIPS, 1.4 for PACS, and 1.3, 1.25, and 1.2 for SPIRE, which not only account for the 
portion of the PSF extending outside the aperture, but also flux contamination from the low-level PSF wing in 
the background annulus. Background emission was calculated as the sigma-clipped mean of an annulus spanning 
1.75 to 2 times the source radius. Flux uncertainties were increased where large gradients existed across the
background annulus and in regions of crowding and confusion.  Photometry was verified quantitatively through 
comparison to SAGE PSF-fit photometry and by manually measuring $\sim$20 sources (including those whose SEDs 
are highlighted below).  No significant systematic offsets were found, and random offsets are consistent with 
the quoted uncertainties.  We find that in these early data we can reliably extract point sources
as faint as (200,300,150,70,40) mJy at (100,160,250,350,500) $\mu$m.  Greater
integration time and improved artifact mitigation will result in fainter
values for the complete HERITAGE survey. The uncertainty in the absolute flux calibration 
is 20\% for PACS \citep{poglitsch10} and 15\% for SPIRE \citep{swinyar2010}.

We  selected a subset of reliable {\it Herschel} sources, most likely YSO candidates,  from 
the more complete source list using the following criteria.  We carefully examined the 
environment of all candidates at all wavelengths, simultaneously with their SEDs, 
aperture photometry, and existing W08 and GC09 catalog photometry.  
Sources were removed that could not be unambiguously identified in
images over a wide wavelength range due to multiplicity or complex
diffuse emission.  We also required reliable photometry spanning at
least 5.8--100 $\mu$m.  This examination is subjective but consistent with
the goals of assessing how {\it Herschel} changes our understanding of star
formation.  Analysis with quantitative completeness limits will be
performed when the higher quality HERITAGE survey data are available.

We removed known non-YSOs from our list.  At {\it Herschel}
wavelengths, background galaxies are the main contaminant as their
far-IR emission arises from star-forming regions. 
We excluded 2 AGN candidates from
\citet{kozlowski09}, as well as 7 probable galaxies based on spatial
morphology in high resolution IRAC images.  
Evolved stellar envelopes lack cold dust, peak in the mid-IR, and
are few among {\it Herschel} sources. We removed one from \citet{boyer10}, 
leaving 207 sources in
our very conservative list.


\begin{figure}
 \centering
 \resizebox{\hsize}{!}{\includegraphics[width=\textwidth]{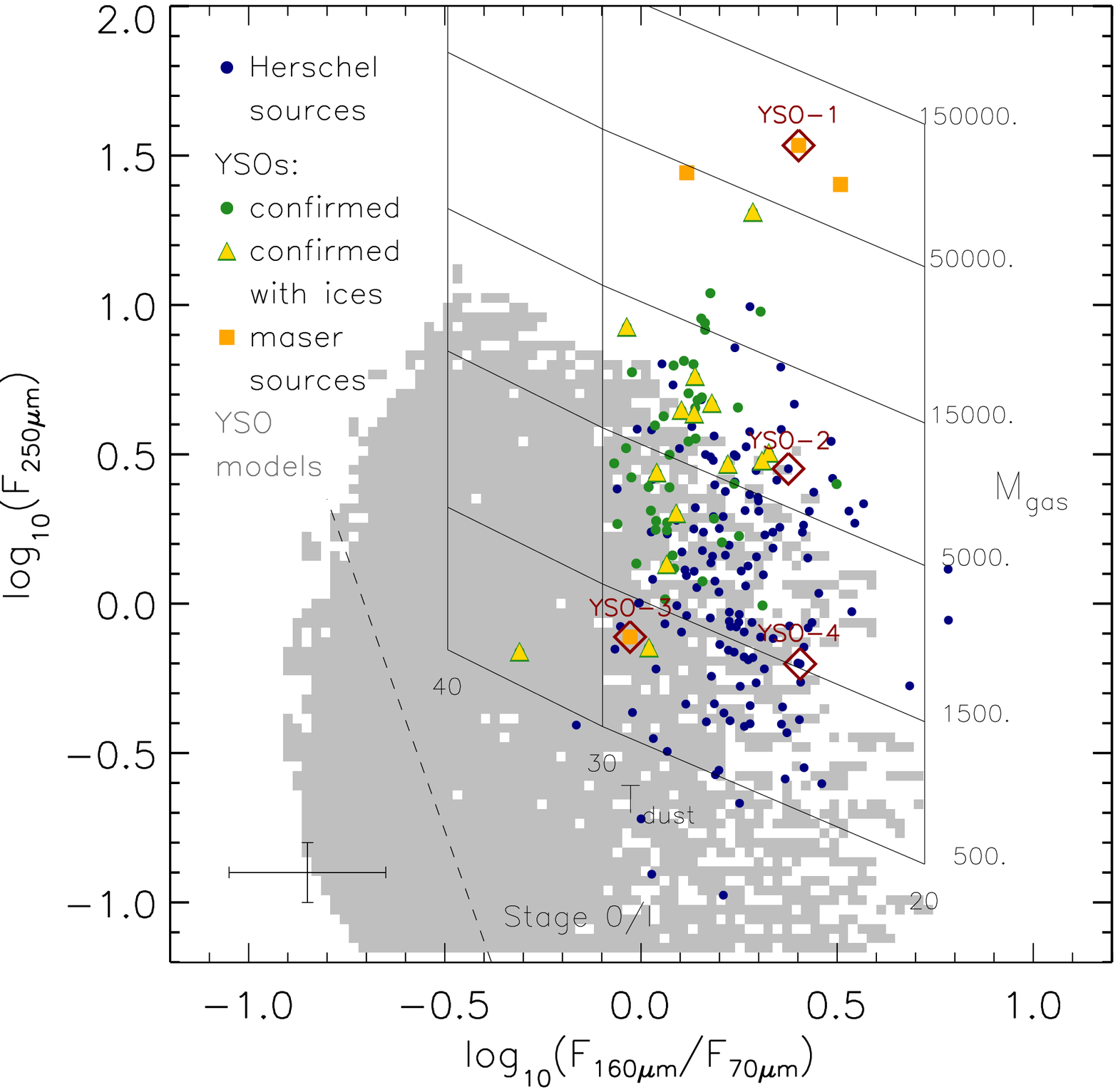}} 
 \caption{Color-magnitude diagram showing F$_{PACS160 \mu \rm m}$/F$_{MIPS70 \mu \rm m}$ vs. F$_{SPIRE250 \mu \rm m}$ [Jy].  
The distribution of YSOs is compared to predictions from YSO models shown in grey (R06); those to the right of the dashed 
line are Stage 0/I models.  We also compare to optically thin greybody emission with M$_{gas}$ = (0.5--150) $\times$ 10$^{3}$ 
M$_\odot$ at 20--40$\;$K, showing that these sources have very cool and massive circumstellar envelopes. 
The {\it Herschel} sources discussed in Sect.~\ref{sedsection} are indicated by red diamonds. A typical error bar is shown in the lower left corner. \label{cmdfig}}
\end{figure}

\section{Far-IR photometric properties of LMC YSOs}

\begin{figure}
\centering
\resizebox{\hsize}{!}{\includegraphics[width=\textwidth]{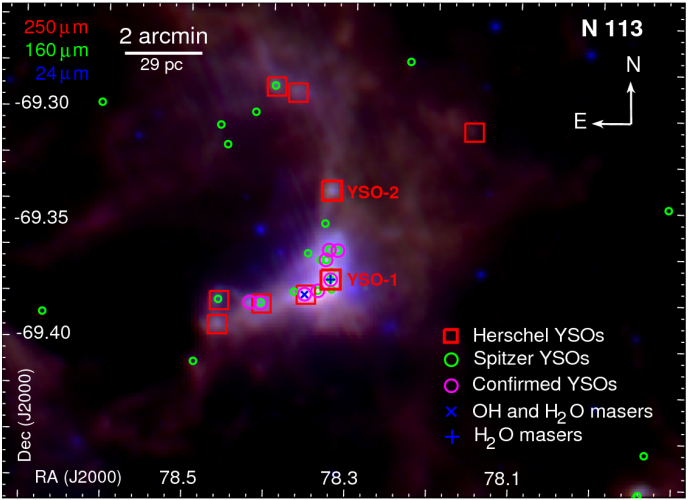}}
 \caption{Three-color composite image of the H\,{\sc ii}  region N\,113 combining SPIRE 250~$\mu$m (red), PACS 160~$\mu$m 
(green),  and MIPS 24~$\mu$m images.  Our reliable {\it Herschel} sources are marked with red boxes.  
{\it Spitzer}-identified YSO candidates, YSOs \textquotedblleft confirmed\textquotedblright\ via spectroscopy or associated with OH and H$_{2}$O masers (\textquotedblleft confirmed\textquotedblright\
YSOs) are also indicated.  Maser positions are from \citet{green08} and \citet{ellingsen10}.  We discuss the SEDs of YSO-1 
and YSO-2 in Sect.\ref{sedsection}. \label{n113fig}}
\end{figure}

Figure~\ref{cmdfig} shows a color-magnitude diagram of our 207  YSO candidates, combining {\it Spitzer} (70 ~$\mu$m) 
with {\it Herschel} (160 and 250~$\mu$m) bands.  Other color combinations show similar source distributions
and lead to similar conclusions.
YSO candidates selected based on photometric data alone require spectroscopic confirmation.  The presence of ice 
absorption is quite definitive; silicate absorption and aromatic emission are at least strongly suggestive (in 
the parsec-sized beam). 
To date, $\sim$300 sources in the LMC have been spectroscopically confirmed or supported \citep{shimonishi08, 
oliveira09, seale09,vanloon10lmc}.  Of these, 126 lie in the HERITAGE SDP area; our list includes 58 (the remainder 
are too faint or or too confused).  We label these as ``confirmed YSOs.''  Photometric YSO candidates may also be 
\textquotedblleft confirmed\textquotedblright\ through maser identification.  In the HERITAGE SDP area, maser emission has been detected in the N\,113 
\citep[H$_{2}$O and OH;][see Fig.~\ref{n113fig}]{whiteoak86,lazendic02,oliveira06,brooks97} and N\,105 \citep[H$_{2}$O, 
OH and methanol;][]{sinclair92,lazendic02,oliveira06} star formation regions.  Each of these regions has two clumps 
of masers, and all four sites are associated with mid-IR sources identified as YSO candidates. Three of these sources 
have been spectroscopically observed (Seale et al. 2009).  

We fit the SED of each source with a large grid of dust radiative transfer models of individual massive YSOs 
(\citealt{r06, r07}, hereafter R06, R07), constraining the circumstellar dust distribution and other physical 
parameters (Sect.~\ref{sedsection}).  Comparison between observed far-IR fluxes and those predicted by the R06 YSO 
model grid confirms the youthful nature of these sources.  All {\it Herschel} sources, and all models to the right (redward) 
of the dashed line in Fig.~\ref{cmdfig} are consistent with Stage~0/I , i.e. the circumstellar mass exceeds that of 
the central source.  In fact, many measured fluxes are redder and brighter than the R06 grid.  We model example sources 
in detail below, and in general this color difference is resolved by adding a larger outer envelope of cool dust.

We also estimate the mass of cool circumstellar dust using simple greybody emission from optically thin dust at a 
single temperature.  A temperature -- mass grid is overplotted on Fig.~\ref{cmdfig}.  We use the same opacity 
curve as the YSO radiative transfer models, i.e. the opacity power-law index $\beta$ is interstellar 
($F_{\lambda} \propto \lambda^{-\beta}$ with $\beta \approx$ 2) for wavelengths $\gtrsim$ 20$\;\mu$m 
\citep{whitney2003a}.


\section{{\it Herschel} view of YSOs and their environments}
\label{n113section}


\onlfig{3}{
\begin{figure*}
\centering
\includegraphics[width=0.49\textwidth]{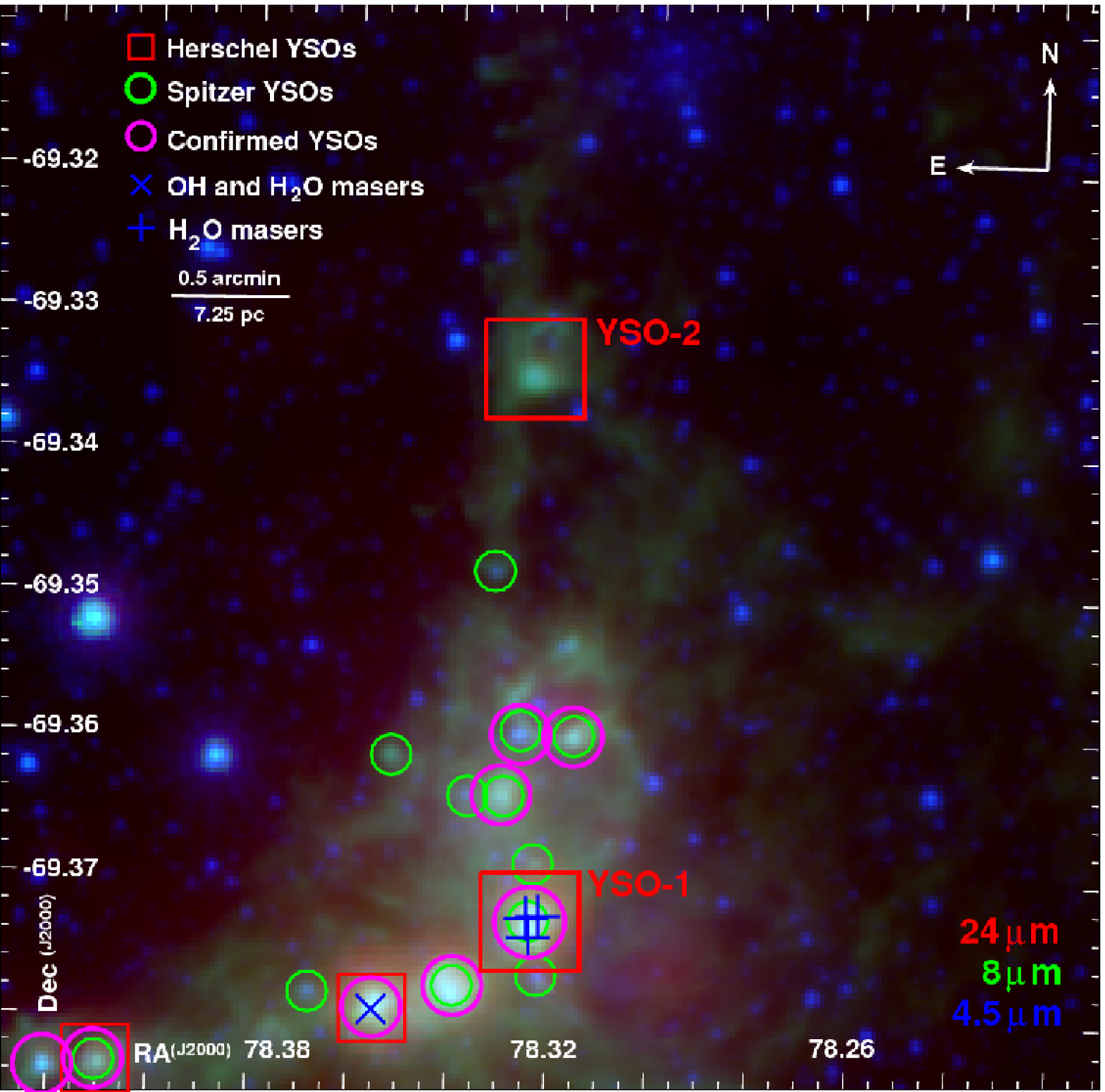} 
\hfill
\includegraphics[width=0.49\textwidth]{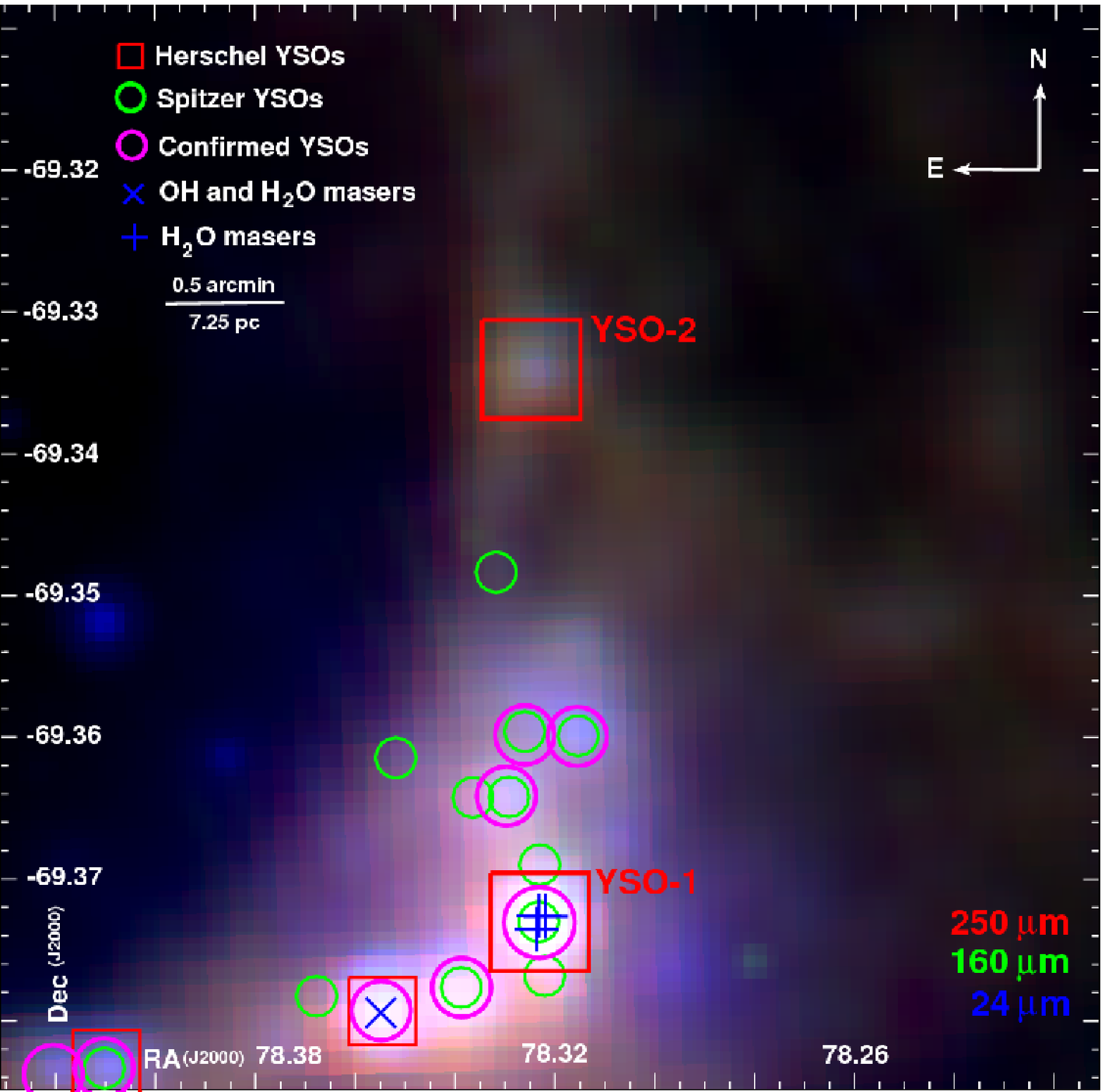} 
\caption{Three-color composite images showing the environment of sources YSO-1 and YSO-2 in the N\,113 massive star formation 
region based on the {\it Spitzer} and {\it Herschel}  observations.  The left-hand image combines IRAC and MIPS, showing 
emission from MIPS 24~$\mu$m, IRAC 8.0~$\mu$m, and IRAC~4.5~$\mu$m (red, green, and blue, respectively).  The image at the 
right combines SPIRE 250~$\mu$m (red), PACS 160~$\mu$m (green), and MIPS 24~$\mu$m (blue) images.  All images are scaled 
logarithmically. The linear distance scale is shown for the LMC distance of 50~kpc \citep{schaefer2008}.  Several categories 
of confirmed YSOs (see text for explanation and references) are marked as indicated in the legend.  The OH and H$_{2}$O 
masers marked by a single cross ($\times$) symbol coincide in position within the uncertainties.  Sources YSO-1 and YSO-2 
are discussed in Sects.~\ref{n113section} and \ref{sedsection}.  Maser positions are from \citet{green08} and 
\citet{ellingsen10}.  YSO-1 was identified as a YSO candidate by \citet[051317.69-692225.0]{gc09} based on the {\it Spitzer} 
data and confirmed spectroscopically by \citet{seale09}, although no ice features were detected.  YSO-2 is a new 
{\it Herschel} YSO candidate.
\label{coloryso1yso2}}
\end{figure*}
}  


\onlfig{4}{
\begin{figure*}
\centering
\includegraphics[width=0.49\textwidth]{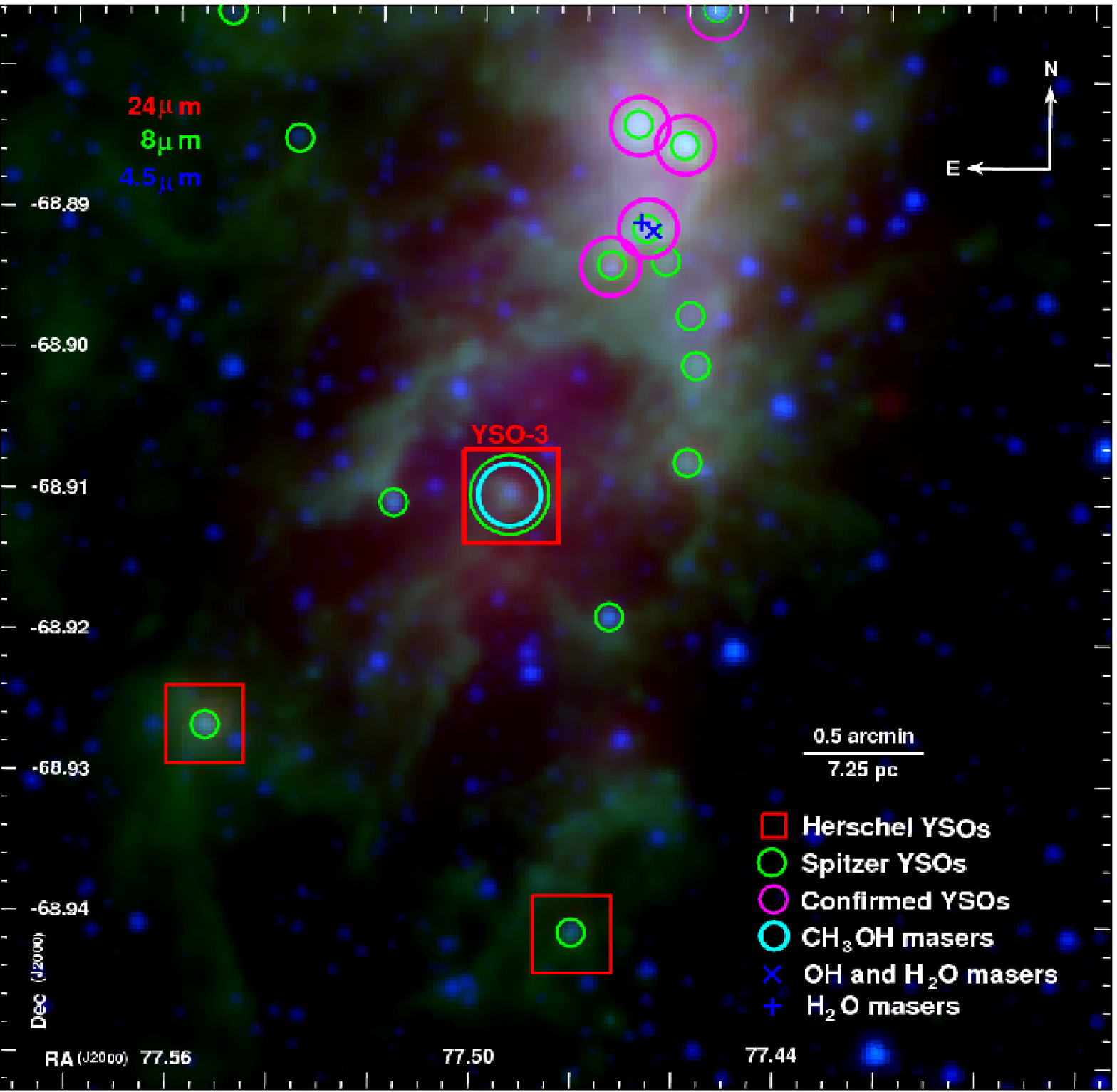}
\hfill
\includegraphics[width=0.49\textwidth]{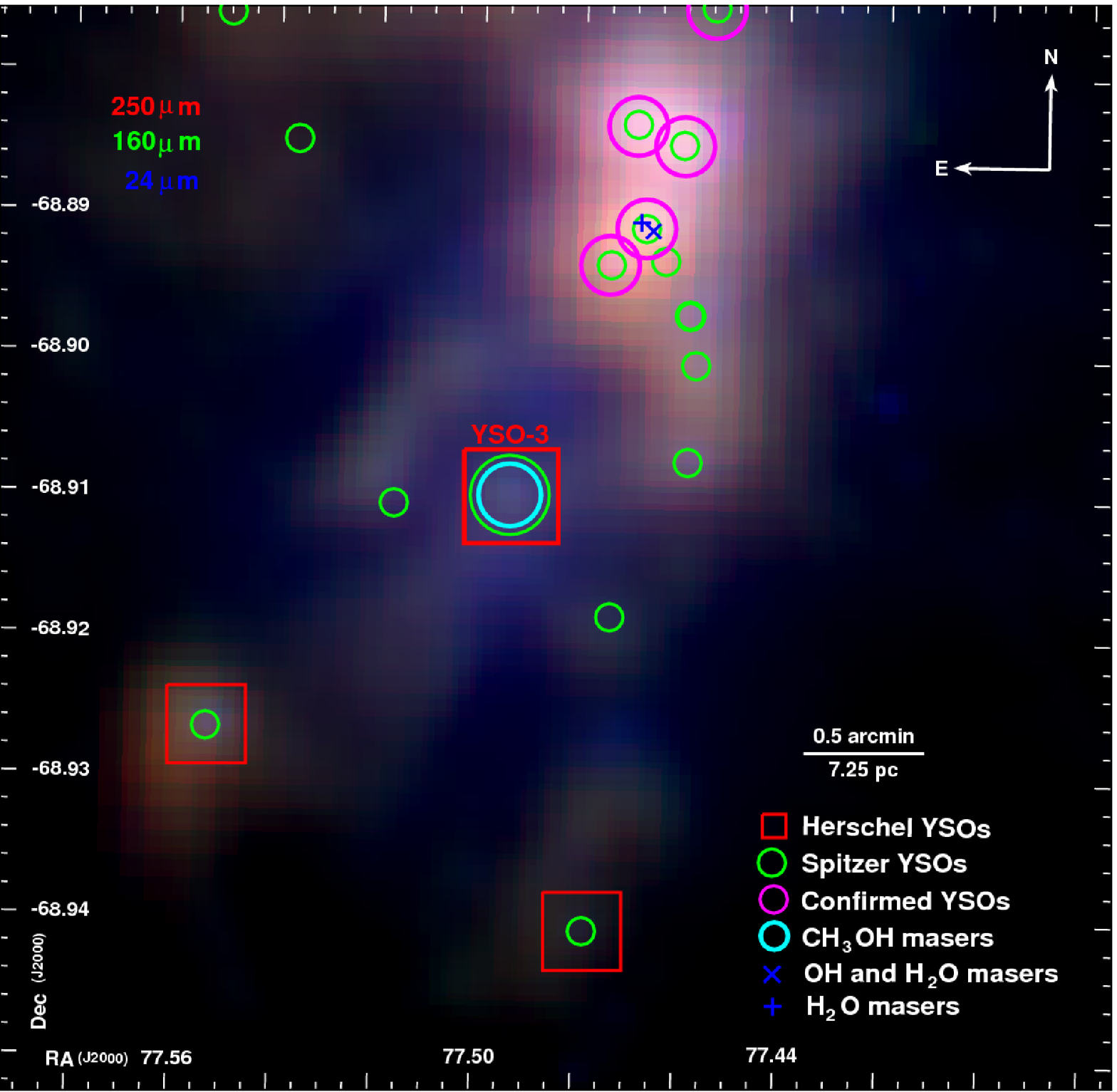}
\caption{Three-color composite images showing the environment of YSO-3 in the LMC N\,105 massive star formation region based 
on the {\it Spitzer} and {\it Herschel} observations.  Coloring and scaling are the same as in Fig.~\ref{coloryso1yso2}.  YSO-3 was identified as a YSO candidate by \citet[050958.52-685435.5]{gc09} based on the 
{\it Spitzer} data. This source is associated with 6.7~GHz and 12.2~GHz methanol masers  
\citep{sinclair92,ellingsen10}. 
\label{coloryso3}} 
\end{figure*}
}  


\onlfig{5}{
\begin{figure*}
\centering
\includegraphics[width=0.49\textwidth]{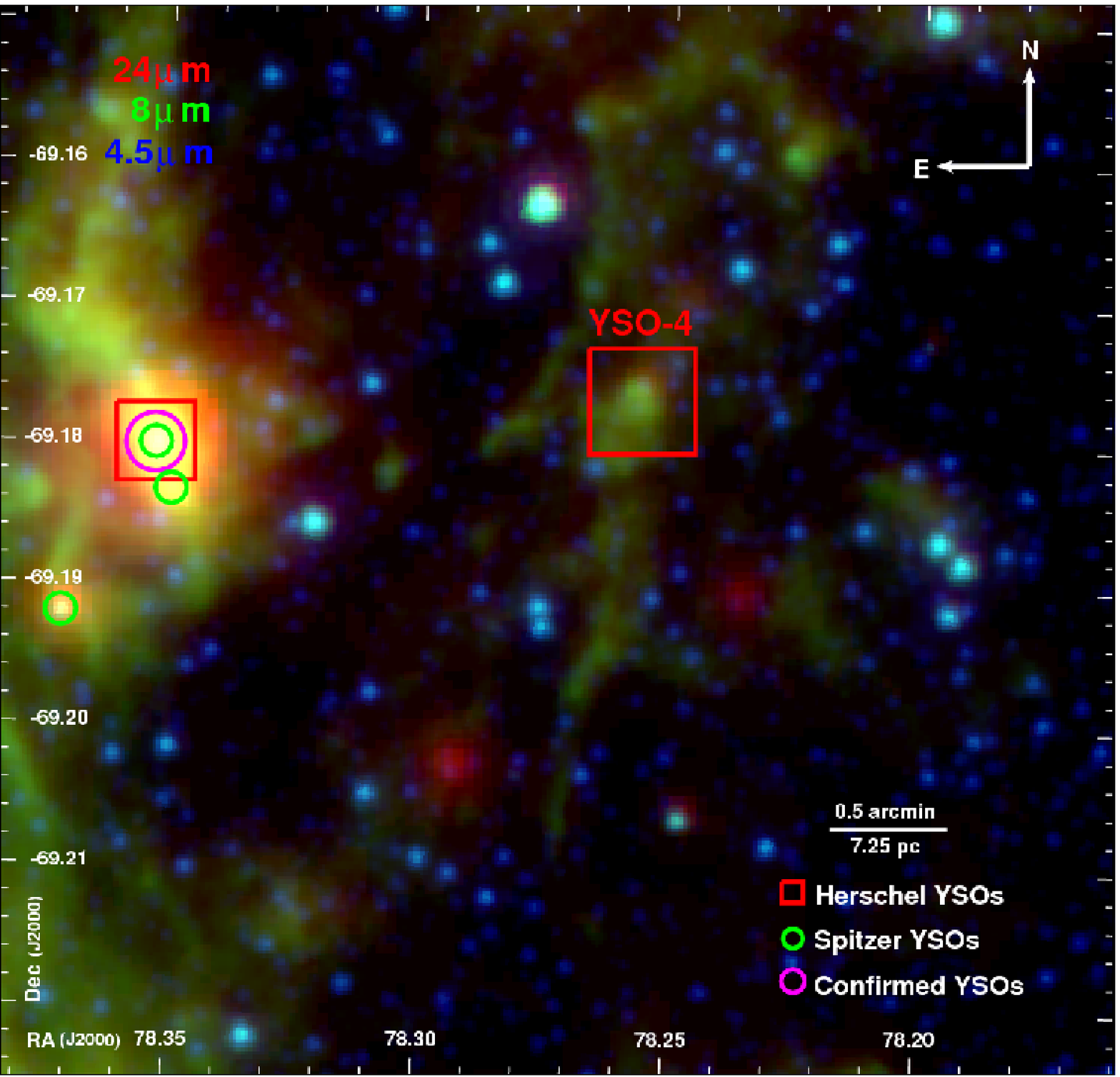}
\hfill
\includegraphics[width=0.49\textwidth]{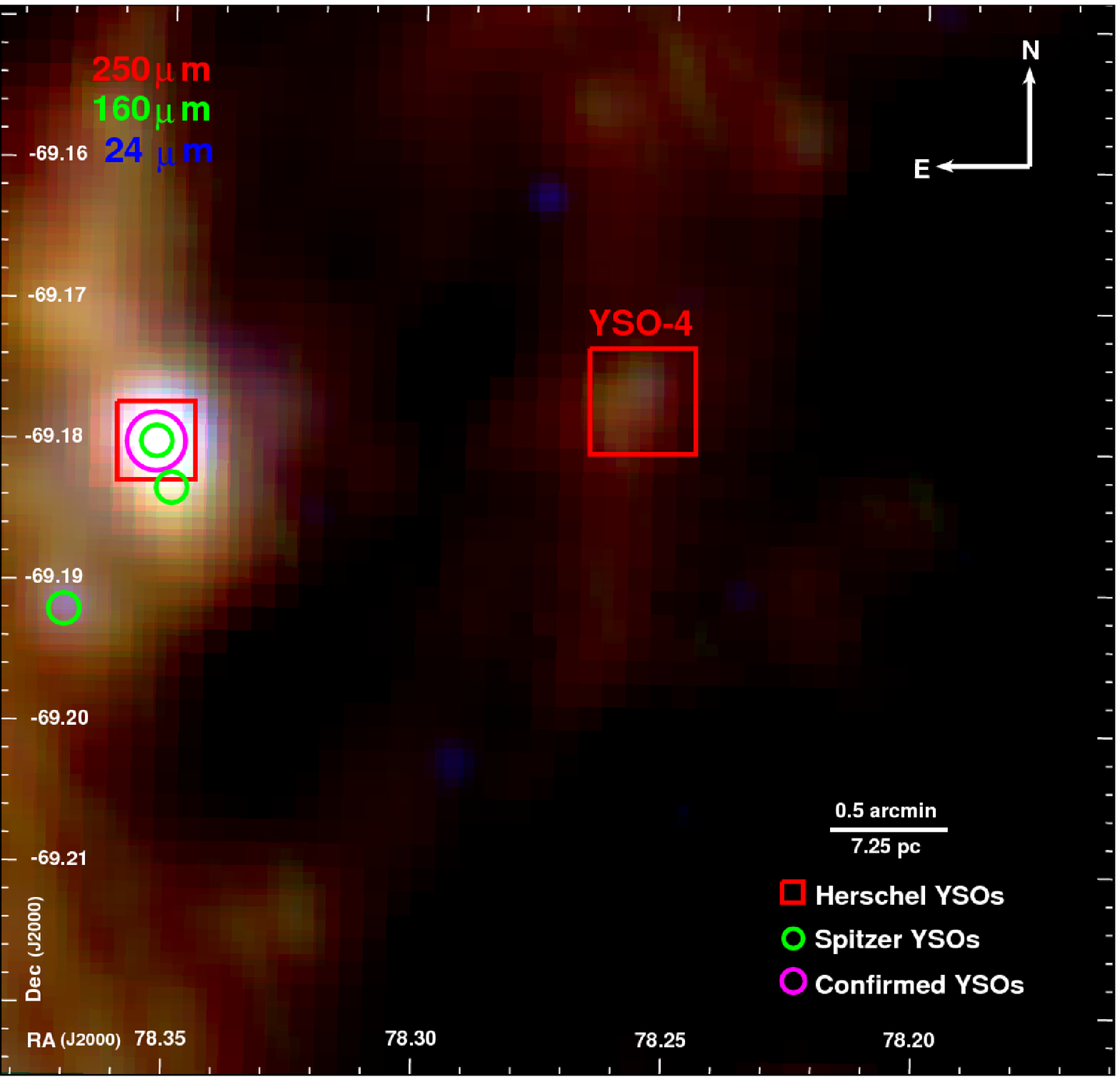}
\caption{Three-color composite images showing the environment of YSO-4 based on the {\it Spitzer} and {\it Herschel}  
observations.  Coloring and scaling are the same as in Fig.~\ref{coloryso1yso2}. YSO-4 is 
located at the edge of an evacuated supershell in the LMC bar, between N\,105 and N\,113.  It does not appear as a 
resolved source in 2MASS imaging and is faint in IRAC bands, excluding it from {\it Spitzer} YSO lists requiring shorter 
wavelengths.  The SED of this evident {\it Herschel} source is quite steep from 24 $\mu$m to {\it Herschel} wavelengths (Fig.~\ref{sedfig}), 
a trend we note in $\sim 20\%$ of the newly-identified {\it Herschel} YSO candidates  \label{coloryso4}}
\end{figure*}
}  

We highlight four YSOs, showing the different ways that {\it Herschel}
can impact star formation studies (SEDs are modeled in
Sect.~\ref{sedsection}).  Two sources are in the N\,113 H\,{\sc ii}
region,
one of the brightest regions in the HERITAGE SDP strip.  Figure~\ref{n113fig} is a three-color composite image of 
N\,113 combining SPIRE 250~$\mu$m, PACS 160~$\mu$m, and MIPS 24~$\mu$m images.  We detect 9 reliable {\it Herschel} sources in N\,113, 
including 4 not previously identified as YSO candidates.    N\,113 presents a clear example of star formation triggered by the 
winds from massive stars \citep{oliveira06}.  Current star formation activity is concentrated in the central lane of remnant 
molecular gas and dust, compressed by a complex structure of ionized gas bubbles, created by massive stars in several young 
clusters ($<10$~Myr).  Ongoing star formation in the lane is pinpointed in its earliest stages by maser emission and by 
continuum emission as massive YSOs evolve \citep{brooks97,wong06}.
Two bright {\it Herschel} sources associated with masers and a third,
fainter source between them are all confirmed YSOs.  We model the full
SED (1.25--500~$\mu$m) of the westernmost of these (YSO-1), to show
{\it Herschel}'s impact on parameters of a known YSO.  We also model a newly
identified {\it Herschel} source YSO-2 that lies in a filament $\sim$4$'$
to the north (Fig.~\ref{coloryso1yso2}, online only).

We present detailed analysis of two other sources, YSO-3 and YSO-4. YSO-3, which appears to lie in a dense knot at the 
rim of a cavity in the molecular cloud of N\,105, is associated with 6.67 GHz and 12.2 GHz methanol maser emission 
\citep[Fig.~\ref{coloryso3}, online only;][]{sinclair92, ellingsen10}.  Methanol masers are closely associated with the earliest stages of massive star formation 
and are powerful probes of young massive stars and protostars. We expect the SED shape of YSO-3 to be representative 
of SEDs of the youngest sources.  YSO-4  is a new {\it Herschel} YSO candidate. It lies approximately 2$\arcmin$ from the nearest 
{\it Spitzer} YSO candidates on the edge of an evacuated supershell in the LMC bar.  The source is very faint shortward of IRAC 
5.8 $\mu$m and then brightens in a \textquotedblleft pillar of creation\textquotedblright\ morphology, indicative of a highly 
embedded YSO (Fig.~\ref{coloryso4}, online only).

\section{Modeled physical properties}
\label{sedsection}

\begin{figure}
\resizebox{\hsize}{!}{\includegraphics[width=\textwidth]{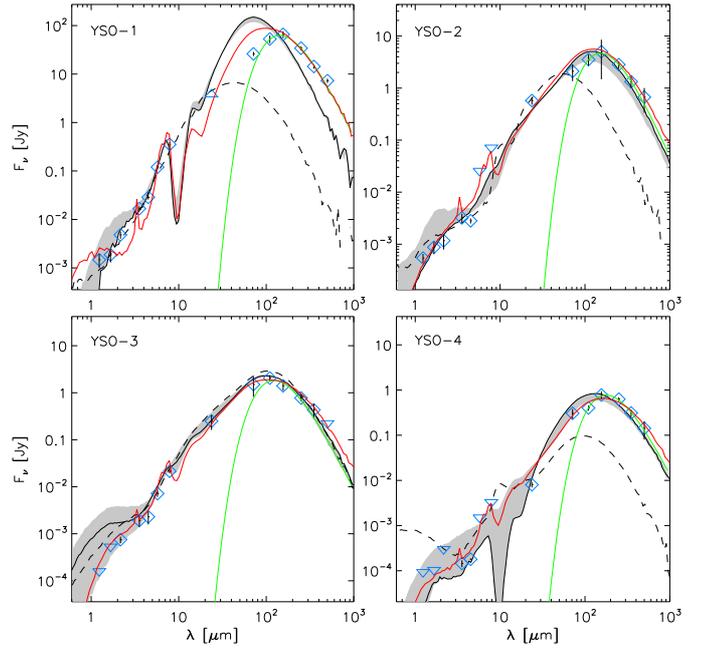}} 
\caption{SEDs of four typical YSOs. Cyan diamonds are the flux values from Table~\ref{datatable}. Cyan triangles 
indicate flux upper limits. The range of well-fitting models from the R06 YSO grid are shown in grey with a black line 
for the single best-fitting model.  The best-fitting R06 model without {\it Herschel} ($\lambda<$50 $\mu$m) is a dashed black 
line.  A single-temperature greybody is shown fit long-wavelengths (green line), and a manually tailored radiative 
transfer model is shown as a red line.  Model parameters are listed in Table~\ref{fittable}.\label{sedfig} }
\end{figure}

Far-IR measurements are critical for deriving the total mass of circum(proto)stellar dust associated with a YSO, and inferring its evolutionary state.  Figure~\ref{sedfig} shows typical sources that illustrate our key findings (photometric data are listed in Table~\ref{datatable}, online only).  
We fit each source with the R06 YSO model grid, first using only pre-{\it Herschel} photometry ($\lambda<$ 50 $\mu$m), and again using all data ($\lambda\leq$ 500 $\mu$m).  Physical parameters derived from these fits are listed in Table~\ref{fittable}.  The addition of {\it Herschel} data provides a much tighter constraint on SED fits, and the range of well-fit models and  uncertainties of derived parameters (as listed in the table) are thus significantly reduced.


\onltab{1}{
\begin{landscape}
\begin{table}
\caption{\label{datatable} {\it Spitzer} and {\it Herschel} fluxes for YSOs discussed in Sect.~\ref{sedsection}.}
\centering
\begin{tabular}{lccccccccccccccccl}
\hline\hline
Source & RA\tablefootmark{a} & Dec & \multicolumn{14}{c}{Fluxes in mJy} & {\it Spitzer} ID (Ref.) \\
\cline{4-16}
      & (h~m~s) & ($^{\circ}$  $'$  $''$) & $F_{J}$ &  $F_{H}$ &  $F_{K}$ &  $F_{3.6 \mu m}$ &   $F_{4.5 \mu m}$ &  $F_{5.8 \mu m}$ & $F_{8.0 \mu m}$ &  $F_{MIPS24 \mu m}$ &   $F_{MIPS70 \mu m}$ &   $F_{PACS100 \mu m}$ &  $F_{PACS160 \mu m}$ &  $F_{SPIRE250 \mu m}$ & $F_{SPIRE350 \mu m}$ &  $F_{SPIRE500 \mu m}$  & Names\\ 
\hline
YSO-1 & 05:13:17.69 & -69:22:25.0 & 1.5(0.5) & 1.9(0.6) & 4.8(0.8) & 17(2)     & 29(3)      & 120(12) & 354(35) & $<$3400  & 26000(2600) &  53000(9600) & 66000(11000) & 34000(3400) & 14000(1400) & 7300(730) & 051317.69-692225.0 (1,2)\\
YSO-2 & 05:13:17.99 & -69:20:04.9 & 0.5(0.1) & 0.9(0.2) & 1.2(0.4) & 3.2(0.7)  & 2.8(0.3)   & $<$24   & $<$68   & 580(96)  & 2100(750)   &  3500(850)   & 5000(3500)   & 2800(460)   & 1300(330)   & 670(340)  & ... \\
YSO-3 & 05:09:58.52 & -68:54:35.5 & $<$0.2   & $<$0.5   & 0.8(0.1) & 1.9(0.5)  & 2.3(0.7)   & 7(1)    & 21(2)   & 250(88)  & 1500(680)   &  2100(300)   & 1400(300)    & 770(110)    & 430(150)    & $<$210    & 050958.52-685435.5 (1) \\
YSO-4 & 05:13:01.55 & -69:10:34.8 & $<$0.1   & $<$0.1   & $<$0.3   & 0.14(0.03)& 0.18(0.01)  & $<$1    & $<$3    & 8(1)     & 310(73)     &  400(51)     & 780(160)     & 630(63)     & 310(55)     & 140(61)   &  ... \\
\hline
\end{tabular}
\tablefoottext{a}{Positions are determined from the highest-resolution images available (IRAC); few-arcsecond offsets present in this early {\it Herschel} data especially at PACS100 are accounted for in the photometry.}
\tablebib{
(1)~\citet{gc09}; (2) \citet{seale09}.
}
\end{table}
\end{landscape}
}

Nearly all sources detected by {\it Herschel} are Stage~0/I, i.e., very young.  A protostar's evolutionary state is often quantified 
by the ratio of circumstellar to stellar mass.  In our single-source YSO models, the accreting flattened envelope has an 
analytic mass distribution \citep[derived by][]{TSC}, and infall rates are proportional to envelope mass within an enclosed 
radius. One caveat is that we do not know if all mass enclosed by the {\it Herschel} beam will fall on to the central 
source, so the derived infall rates may be upper limits.  We find that the accretion rate relative to the central mass, 
$\dot{M}_{env}/M_{\star}$ is a good measure of evolutionary state (e.g.,\citealt{whitney2003b}; \citealt{ri_m16}; R06; R07).
Without {\it Herschel} data, $\dot{M}_{env}/M_{\star}$ of our sources is (2,5,1,4)$\times$10$^{-5}\;$yr$^{-1}$
for (YSO-1,YSO-2,YSO-3,YSO-4).  When
{\it Herschel} photometry is included, $\dot{M}_{env}/M_{\star}$ increases to (2,3,2,4)$\times$10$^{-4}\;$yr$^{-1}$. 
The higher infall rates are in agreement with previous outflow studies of massive protostars \citep{churchwell1999}.
The implied ages of $<$10$^4$ yr further strengthen the assertion that {\it Herschel} probes the very youngest sources.


\begin{table}
\setlength{\tabcolsep}{3pt}
\centering
\caption{\label{fittable} Modeled YSO parameters, fitting each source
using the R06 YSO grid and only pre-{\it Herschel} data ($\lambda<$ 50 $\mu$m),
the R06 grid and all data ($\lambda\leq$ 500 $\mu$m),  and optically 
thin single-temperature dust emission using data $\lambda>$ 100 $\mu$m (greybody).}
\begin{tabular}{ccccccc}
\hline\hline
 &  & $\lambda$  &log(L$_{bol}$) & log($\dot{\rm M}_{env}$) & log(M$_{disk}$) & M$_{\star}$\\
\raisebox{1.5ex}[0pt]{Source}&\raisebox{1.4ex}[0pt]{Model}&($\mu$m)& (L$_\odot$) & (M$_\odot\;$yr$^{-1}$) & (M$_\odot$) & (M$_\odot$)\\
\hline
YSO-1     & R06 & $<$50     & 5.1$\pm$0.2 & -3.3$\pm$0.8 &  -1.0$\pm$1.1 & 27$\pm$5\\
          & R06 & $\leq$500   & 5.5$\pm$0.1 & -2.1$\pm$0.0 &  -0.1$\pm$0.4 & 45$\pm$5\\ [2mm]
YSO-2     & R06 & $<$50     & 4.5$\pm$0.5 & -3.2$\pm$0.9 &  -0.9$\pm$0.7 & 13$\pm$2\\
          & R06 &  $\leq$500   & 4.5$\pm$0.3 & -2.4$\pm$0.2 &  -0.7$\pm$0.7 &13$\pm$2 \\ [2mm]
YSO-3     & R06 & $<$50     & 4.2$\pm$0.2 & -3.8$\pm$1.8 &  -1.2$\pm$1.1 & 12$\pm$1\\
          & R06 &  $\leq$500   & 3.9$\pm$0.2 & -2.5$\pm$0.2 &  -0.8$\pm$0.8 & 13$\pm$1\\ [2mm]
YSO-4     & R06 &  $<$50     & 3.3$\pm$0.5 & -3.6$\pm$1.1 &  -1.4$\pm$0.9 & 7$\pm$1\\
          & R06 & $\leq$500   & 3.3$\pm$0.2 & -2.5$\pm$0.1 &  -0.8$\pm$0.7 & 9$\pm$1\\
\hline
&  & $\lambda$    & log(L$_{far-IR}$) & T$_{dust}$   & log(M$_{gas}$)  \\
 \raisebox{1.5ex}[0pt]{Source}&  \raisebox{1.4ex}[0pt]{Model} & ($\mu$m) & (L$_\odot$)     & (K) & (M$_\odot$)  \\
\hline
YSO-1  &  greybody &$>$100  & 5.0 & 20 & 4.7  \\
YSO-2  &  greybody &$>$100  & 3.9 & 19 & 3.8 \\
YSO-3  &  greybody &$>$100  & 3.6 & 24 & 2.9  \\
YSO-4  &  greybody &$>$100  & 3.0 & 16 & 3.3  \\
\hline
\end{tabular}
%
\end{table}

Observationally, the ratio of far-IR/submm luminosity to bolometric  luminosity is often used as a proxy 
for the circumstellar to total mass ratio. \citet{andre93} define Class~0 as having L$_{\lambda>350\mu{\rm
m}}$/L$_{bol}>$5$\times$10$^{-3}$.
We calculate a similarly simple evolutionary measure, fitting an optically thin greybody to the data $\lambda>$100 $\mu$m, and find that it agrees reasonably with the YSO model fit results (Model \textquotedblleft greybody\textquotedblright\ in Table~\ref{fittable} and green line in Fig.~\ref{sedfig}).  Resultant dust mass and far-IR/submm luminosity depend on the dust opacity power-law index. We use $\beta$ = 2; a different choice does not change the conclusion that these sources are highly embedded. We find L$_{far-IR}/$L$_{bol} \sim 0.2-0.5$ for our four examples, very high and consistent with the \citet{andre93} Class~0 definition.  We also find T$_{dust}\sim16-24$~K, consistent with the far-IR colors (Fig.~\ref{cmdfig}) and expectations for the least evolved YSOs.

Even the best-fitting R06 models (dashed line in Fig.~\ref{sedfig}) often fall well below the data at longer 
wavelengths (also apparent in Fig.~\ref{cmdfig}).  The R06 models were designed for analysis of {\it Spitzer} data
 at shorter wavelengths; models have outer envelope radius R$_{out}\leq 0.5$~pc, T$ \geq 30$~K for 
luminous sources.  The envelope of cooler dust and gas farther from the source emits  
little at $\lambda\lesssim$100 $\mu$m but can contribute significantly in the {\it Herschel} bands.  

We created new dust radiative transfer models (red line on Fig.~\ref{sedfig}) and find that implied accretion rate and other 
{\em critical evolutionary  parameters do not change significantly} from the R06 best fits.  
The new models have larger envelope radii, heating by external radiation, and PAH emission.
We used the R06 best-fit models as starting points. $R_{out}$ was varied 
between the R06 maximum of 0.5 pc and 5 pc, with best-fitting 
values of (3.4,1,1,1) pc. The importance of external heating was assessed by comparing models with
none to models with a \citet{mathis1983} diffuse interstellar radiation
field extincted by A$_V$=1. In all cases the latter improved the fits.
Any increase in the outer envelope radius also increases dust absorption
at short wavelengths, the effect of which was balanced by varying the density
in the outflow cavity between 0.1 and 1 times the R06 best-fit value.
Reduced values of (1,0.1,2,2)$\times$10$^{-20}$g$\;$cm$^{-3}$ provided
the best fits to the data.
The envelope infall rates and source luminosities remained the same, showing that the R06 grid
can still be used to interpret the {\it Herschel} data points.
Adding PAH emission to our models results in better fits to bands effected by aromatic emission (predominantly IRAC 5.8 and 8.0~$\mu$m).  Additional work must also be done to account for blending of multiple sources at Magellanic distances, as YSOs are frequently part of multiple systems and small unresolved clusters.  

Optimal analysis of  Magellanic YSOs observed by {\it Herschel} and {\it Spitzer}
will benefit from improvements to the R06 model grid.  However, it is the
inclusion of {\it Herschel} data, even with the older models, that is most
significant: If these 4 embedded YSOs are typical, as we expect,
the statistics of YSO physical parameters in the Magellanic System will
be significantly clarified.

\section{Conclusions}

We show that {\it Herschel} far-IR photometry is very effective in identifying YSOs in the LMC.  
Adding {\it Herschel} data to existing {\it Spitzer} and near-IR observations results in significantly improved
analysis of YSOs, as summarized in these key findings:
\begin{list}{$\circ$}{\setlength{\topsep}{0in}\setlength{\parsep}{0in}}
\item Nearly all sources detected by {\it Herschel} are Stage~0/I, very young, with a high ratio of circumstellar 
to stellar mass. \item Previously studied warm sources such as YSO-1 require more circumstellar dust to 
fit the {\it Herschel} data, implying a less evolved state than would be inferred from {\it Spitzer} data alone.
\item {\it Herschel} photometry significantly constrains our SED fits, decreasing the range of circumstellar 
dust masses and evolutionary states consistent with the measurements.
\item Many sources require even more cold circumstellar dust than is present in our original model grid, 
motivating improvements to our models.
\end{list}
Our observations prove  that {\it Herschel} can offer us,  for the first time, an inventory of the earliest 
stages of protostellar formation throughout an entire galaxy. \\

\begin{acknowledgements}
    
We acknowledge financial support from the NASA Herschel Science Center, JPL 
contracts \# 1381522 \& 1381650.  We thank the contributions and support from 
the European Space Agency (ESA),  the PACS and SPIRE teams,  the Herschel 
Science Center and the NASA Herschel Science Center  (esp. A. Barbar and K. Xu) 
and the PACS and SPIRE instrument control center at CEA-Saclay,  without which 
none of this work would be possible.

\end{acknowledgements}

\bibliographystyle{aa} 
\bibliography{refs_14688} 

\end{document}